# Preliminary Study on Forced Oscillation of Power System with Quadratic Nonlinearity

Zhou Yichen, *Member, IEEE*, Wu Jianwei

*Abstract*—Forced oscillation (FO) is a significant concern threating the power system stability. Its mechanisms are mostly studied via linear models. However, FO amplitude is increasing, e.g., Nordic and Western American FOs, which can stimulate power system nonlinearity. Hence, this paper incorporates nonlinearity in FO mechanism analysis. The multi-scale technique is employed in solving the forced oscillation equation to handle the quadratic nonlinearity. The amplitude-frequency characteristic curves and first-order approximate expressions are derived. The frequency deviation and jumping phenomenon caused by nonlinearity are discovered and further analyzed by comparing with linear models. This paper provides a preliminary research for nonlinear FOs of power system, and more characteristics should be further analysis in the near future.

## I. INTRODUCTION

Continuous forced oscillations (FO) usually appears when the disturbance frequency is close to the frequency of power system natural mode. Even small disturbances could cause huge power oscillations [1]. Numerous FOs have been detected, such as Nordic, North America, and China [2]. Plenty of resonance disturbances are proved to be causes, e.g., the fluctuation of turbine, load, wind, and controller [3,4].

Many valuable mechanisms of FO have been contributed by researchers. Explicit expressions of FO in multi-machine power systems are derived [5], [6]. An energy-based method is proposed to locate oscillation sources in power systems [7]. Normal Form method is studied and interprets the harmonic oscillation caused by nonlinearity [8], [9]. However, there are still some phenomena that have not been revealed, such as frequency deviation in power system oscillations [10]. Power transmission through a long distance may increase the load rate of transmission line, and leads power systems to operate in the nonlinear areas. Therefore, it's vital to reveal significant characteristics of nonlinear FO for identification and detection.

To tackle this problem, this paper employs a multi-scale method to study the mechanism of nonlinear FO. The characteristic curves of amplitude-frequency and first-order approximate expression are derived to interpret the influence of nonlinear factors. Main contributions of this paper are: 1) The first-order approximate analytic expression of nonlinear FOs are presented. 2) The frequency deviation and jumping phenomenon caused by power system nonlinearity are observed and interpreted.

This article is organized as follows. Section II describes the nonlinear model. Section III details the application of a multi-scale method in nonlinear FO analysis, and approximates resonance solution. Section IV interprets the resonance curve and frequency deviation of nonlinear FO by analytic expression and simulations. Section V gives the conclusion.

## II. NONLINEAR MODEL OF SMIB

A single-machine infinite-bus (SMIB) power system model with periodic loads is established and shown in Fig. 1 to analyze the nonlinear FO. The SMIB mathematical model is written as:

$$\begin{cases} 2H \dfrac{d\Delta\omega}{dt} = P_m - P_e - D\Delta\omega + P_d \cos(\omega t) \\ \dfrac{d\Delta\delta}{dt} = \omega_{\text{base}} \Delta\omega \end{cases} \quad (1)$$

where, $\delta$, $\omega$, $H$, $D$, $P_m$ and $P_e$ are the rotor angle, rotor speed, inertial time constant, equivalent damping coefficient, input mechanical power, and output electromagnetic power of G1, $\omega_{\text{base}}$ is the reference of power system frequency, $P_d$ and $\Omega$ are the amplitude and frequency of load disturbance. As $P_e$ can be expressed as

$$P_e = \dfrac{V_G V_B}{X_G} \sin\delta = P_{\max} \sin\delta \quad (2)$$

where $V_G$ is the terminal voltage of G1, $V_B$ is the voltage of the infinite bus, $X_G$ includes transient and line reactance.

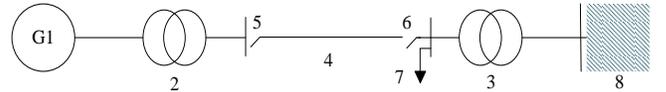

Fig. 1. Single-machine infinite-bus system.

Define the following transformation expressions:

$$\tau = t\sqrt{\dfrac{D}{2H}},\ x(\tau)=\Delta\delta,\ y(\tau)=\sqrt{\dfrac{2H}{D}}\omega_{\text{base}}\Delta\omega,\ a_1 = \dfrac{\omega_{\text{base}} P_m}{D},$$

$$b = \dfrac{\omega_{\text{base}} P_{\max}}{D},\ f = \dfrac{\omega_{\text{base}} P_d}{D},\ \Omega = \omega\sqrt{\dfrac{2H}{D}},\ c = \sqrt{\dfrac{D}{2H}}.$$

By substituting (2) into (1) and adopting the above transformation expressions, (1) is equivalent to

$$\begin{cases} \dfrac{dy}{d\tau} = a_1 - b\sin x - cy + f\cos(\Omega\tau) \\ \dfrac{dx}{d\tau} = y \end{cases} \quad (3)$$

Then, solving the equilibrium point of (3) will yield the result $(x_1,0) = [\arcsin(a_1/b), 0]$. Rewritten (3) in the form of Taylor series expansion at $(x_1,0)$ will generate

$$\ddot{x} + c\dot{x} + \omega_0^2 x - \alpha x^2 = f\cos(\Omega\tau) \quad (4)$$

where $\omega_0^2 = b\cos(x_1)$, $\alpha = b/2\sin(x_1)$, $c$ is the damping of G1 and $f$ is the amplitude of external load excitation.

## III. First-Order Approximate Analytic Expression by Multi-scale Method

To obtain an effective approximate solution, the multi-scale method is employed to reform (4). First, a small parameter $\varepsilon$ is introduced. Then, reform (4) via amplifying $\alpha$ by $\varepsilon$, and amplifying $c$ and $f$ by $\varepsilon^2$, which results in

$$\ddot{x} + \varepsilon^2 c\dot{x} + \Omega^2 x - \varepsilon^2 \sigma x - \varepsilon\alpha x^2 = \varepsilon^2 f\cos(\Omega\tau) \quad (5)$$

where $\sigma$ is the ratio of frequency away from $\omega_0$, and $\Omega^2 = \omega_0^2 + \varepsilon^2\sigma$.

Taking different time scales into consideration, the solution formula can be expressed as

$$x(t,\varepsilon) = x_0(T_0,T_1,T_2) + \varepsilon x_1(T_0,T_1,T_2) + \varepsilon^2 x_2(T_0,T_1,T_2) \quad (6)$$

where $T_0 = t, T_1 = \varepsilon t, T_2 = \varepsilon^2 t$ represent different time scales.

Substitute (6) into (5) and decompose the result into different scales, one can obtain the following differential equations

$$D_0^2 x_0 + \Omega^2 x_0 = 0 \quad (7)$$

$$D_0^2 x_1 + \Omega^2 x_1 = -2D_0 D_1 x_0 + \alpha x_0^2 \quad (8)$$

$$D_0^2 x_2 + \Omega^2 x_2 = -2D_0 D_1 x_1 - D_1^2 x_0 - 2D_0 D_2 x_0 \\ + \sigma x_0 + 2\alpha x_0 x_1 - cD_0 x_0 + f\cos(\Omega\tau) \quad (9)$$

The solution of (7) can be represented as:

$$x_0 = A(T_1,T_2)e^{i\Omega T_0} + \bar{A}(T_1,T_2)e^{-i\Omega T_0} \quad (10)$$

where $\bar{A}$ is the conjugate complex number of $A$.

Based on (10) and (8), the following equation can be obtained

$$D_0^2 x_1 + \Omega^2 x_1 = -2i\Omega D_1 A e^{i\Omega T_0} + \alpha A^2 e^{2i\Omega T_0} + \alpha \bar{A}^2 e^{-2i\Omega T_0} \\ + 2i\Omega D_1 \bar{A} e^{i\Omega T_0} + 2\alpha A\bar{A} \quad (11)$$

To eliminate the duration term, $D_1 A$ must equal to zero. Thus, $A(T_1,T_2) = A(T_2)$ and the special solution of the (11) is

$$x_1 = -\frac{\alpha A^2}{3\Omega^2} e^{2i\Omega T_0} + \frac{2\alpha}{\Omega^2} A\bar{A} + cc \quad (12)$$

where $cc$ stands for the complex conjugation of the preceding two terms. Further substituting (10), (12) into (9), the following equation is yielded

$$D_0^2 x_2 + \Omega^2 x_2 = [-2i\Omega D_2 A - ci\Omega + \sigma A + \frac{10\alpha^2}{3\Omega^2} A^2 \bar{A} \\ + \frac{1}{2}f]e^{i\Omega T_0} - \frac{2\alpha^2}{3\Omega^2} A^3 e^{3i\Omega T_0} + cc \quad (13)$$

Let the duration terms equal zero, then (14) is obtained

$$-2i\Omega D_2 A - ci A\Omega + \sigma A + \frac{10\alpha^2}{3\Omega^2} A^2 \bar{A} + \frac{1}{2}f = 0 \quad (14)$$

where, define $A(T_2) = \frac{1}{2}ae^{i\varphi}$ and substitute it into (14). Next, the real part equation and imaginary part equation of (14) can be extracted as follows

$$\dot{a} = -\frac{c}{2}a - \frac{f}{2\Omega}\sin\varphi \quad (15)$$

$$\dot{\varphi} = -\frac{5\alpha^2 a^2}{12\Omega^3} - \frac{\sigma}{2\Omega} - \frac{f}{2\Omega a}\cos\varphi \quad (16)$$

With $a$ and $\varphi$ functions of $T_2$. Steady-state solutions of (15) (16) satisfy $f\sin\varphi = -ca\Omega$ and $f\cos\varphi = -5\alpha^2 a^3/6\Omega^2 - \sigma a$. When $\dot{a}=0$, $\dot{\varphi}=0$, the steady-state solution could be obtained from the following equation

$$(ca\Omega)^2 + (-\frac{5\alpha^2 a^3}{6\Omega^2} - \sigma a)^2 = f^2. \quad (17)$$

Based on (17), the steady amplitude of $a$ can be calculated by substituting $f$ and $\sigma$. Then, characteristic curves of amplitude-frequency can be determined by solving the function $a(f)$.

## IV. Analysis and Simulations

This section adopts the basic SIMB parameters $P_m = 5 p.u$, $P_{\max} = 0.6 p.u$, $D = 4 p.u$ and $H = 4s$. Based on small signal stability analysis, the electromechanical mode is solved to be -0.25+j3.60, whose natural frequency is 0.57Hz and damping ratio is 0.07. Set the parameter in (4) as $\alpha = 19.63$, $c = 0.707$, an $\omega_0 = 5.5830$. Employing this test system, characteristics of nonlinear oscillations and forced oscillation curves are studied.

### A. Characteristics of Nonlinear Oscillations

Firstly, this section reveals differences between nonlinear oscillations and linear oscillations based on amplitude-frequency characteristics. Both linear and nonlinear amplitude-frequency curves are calculated (Fig. 2) under different load disturbances $P_d$=0.005, 0.010, 0.015, 0.020, and 0.025. The nonlinear characteristics are represented by solid lines and the linear characteristics are illustrated by dashed lines.

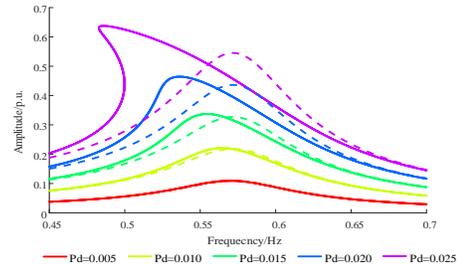

Fig. 2. Amplitude-frequency curves with different $P_d$.

It can be seen from Fig. 1 that for linear oscillations, the maximum amplitude of the forced oscillation occurs at the natural frequency of power system. However, for the nonlinear oscillations, as $P_d$ increases, the frequency of maximum amplitude gradually shifts towards the left, which means that the system frequency gradually departs from the natural frequency of the equilibrium point. Moreover, although all FO amplitudes rise with the excitation amplitude increasing, the amplitude of nonlinear FO grows with the increments of disturbances strength in a nonlinear increase relation.

Then, for further analysis, the impact of different factors on nonlinear FOs are explored in the following contents. Let $\alpha$=0, $0.2\alpha_0$, $0.4\alpha_0$, $0.6\alpha_0$, $0.8\alpha_0$, and $\alpha_0$, respectively, the corresponding amplitude-frequency curves are drawn in Fig. 3. It can be seen that the deviation amount of frequency is increasing faster than the increments of non-linearity degree. Therefore, high attention should be placed on the non-linearity strength of power systems which has an important impact on the frequency deviation.

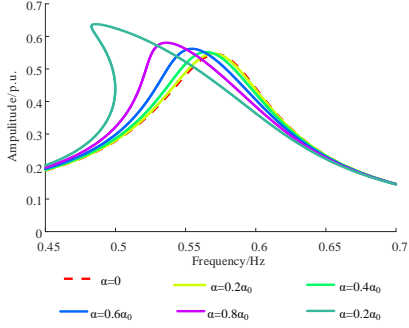

Fig. 3. Amplitude-frequency curves with different $\alpha$.

Set damping to $D$=4, 6, 8, 10, and 12, respectively, the corresponding amplitude-frequency curves are solved in Fig. 4. It shows that as $D$ increases, both the maximum amplitude and the frequency deviation decrease. Hence, a good damp ratio is very helpful to reduce the damage of FOs.

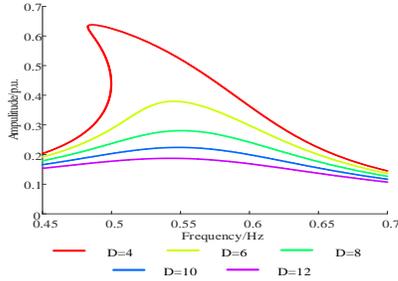

Fig. 4. Amplitude-frequency curves with different damping ratio.

Assign inertia time constant to $H$=2.5, 3, 3.5, and 4, respectively, the corresponding amplitude-frequency curves are exhibited in Fig. 5(a). It can be seen that the maximum amplitude rises as $H$ increases. And the natural frequency and frequency deviation are moving to the left with $H$ increasing. However, as shown in Fig. 5(b), if D/H keeps 1.5 when $H$ increases, the maximum amplitude will decrease. Though increasing $H$ will reduce the impact of load disturbance, it also decreases the damp ratio of power systems. Therefore, when measures are taken to increase equipment virtual inertia, its damage to system damping ratio should also be taken in to consideration.

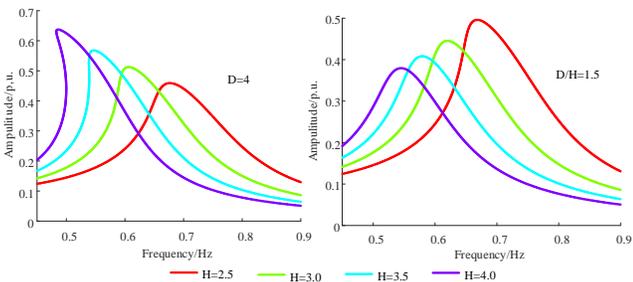

Fig. 5. Amplitude-frequency curves with different inertia.

In addition, the amplitude-frequency curves with different transmission power are shown in Fig. 6. As is shown, with the mechanical power $P_m$ increasing, the maximum amplitude grows at a faster rate than the increments of the power system nonlinearity, and the natural frequency and frequency deviation are moving to the left. Therefore, improving the load rate of the transmission line is another measure to reduce the FO damage.

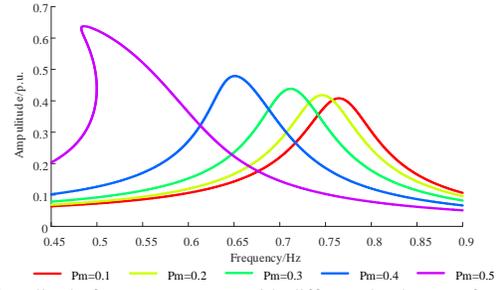

Fig. 6. Amplitude-frequency curves with different load rates of transmission line

In further analysis, a unique jump-phenomenon is found in nonlinear systems and shown in Fig. 7. Assuming that the magnitude of the excitation amplitude remains unchanged. The initial point of the excitation frequency is at point 1. Slowly enlarge the excitation frequency, the vibration amplitude will slowly rise to point 2. Continue to increase the excitation frequency, the amplitude will suddenly soar to point 3, an increase-jump occurs and the subsequent amplitude decreases slowly. If the starting point of the excitation frequency is point 4, gradually decrease excitation frequency then the amplitude will gradually increase to point 7 and then suddenly drop, and a drop-phenomenon occurs. Thus, two amplitudes may appear simultaneously in some areas. Generally, FOs in power systems always increase from small to large, therefore, the steadiest amplitude is the low amplitude curve composed of 1-2-3-4.

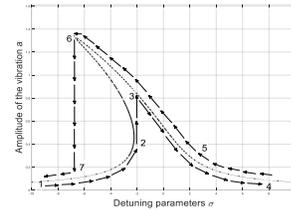

Fig. 7. Jump phenomenon.

### B. Analysis on Forced Oscillation Curve

According to (6), (10) and (12), the first order steady-state approximate solution of the FO curve is:

$$x(t) = Ae^{i\Omega T_0} + \varepsilon(-\frac{\alpha A^2}{3\Omega^2}e^{2i\Omega T_0} + \frac{2\alpha}{\Omega^2}A\bar{A}) + cc \quad (18)$$

where $A = \frac{1}{2}ae^{i\varphi}$, $a$ and $\varphi$ are given by (15) and (16).

As an example, when the frequency of disturbance is 0.55Hz, the FO curves in Fig. 8 (a) can be solved by (18), where simulation results are also provided for comparison.

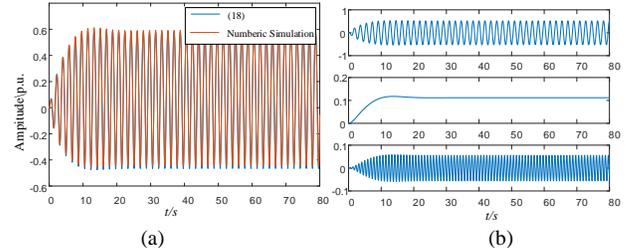

Fig. 8. The curve of (18) in (a), with comparison to numeric simulations in (b).

As shown in Fig. 8(a), the solved FO curves exhibit little difference from numeric simulation results. In addition, components of (18) are given. In Fig. 8 (b), the first figure is the main resonance component, the second one is the DC component, and the last one is the double frequency component. Thus, the simulation curve in Fig. 8 (a) is mainly composed of the main resonance component and the DC components. As for the DC component, it adds a positive shift to the FO curve.

FO curves at different frequencies are further calculated and shown in Fig. 9. As shown in Fig. 9(a), the oscillation amplitude is largest at the 0.52 Hz disturbance, rather than 0.57 Hz disturbance which equals to the system natural frequency. In Fig. 9(b), the maximum amplitude of the amplitude-frequency curve in the jump phenomenon is consistent with FO curves in Fig. 9(a) at 0.52Hz.

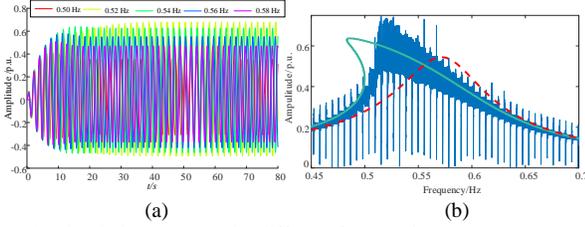

Fig. 9. The simulation curve under different frequencies.

When f=0 and $\sigma=0$, (15) and (16) will be expressed as

$$\dot{a} = -\frac{c}{2}a \quad \dot{\varphi} = -\frac{5\alpha^2 a^2}{12\Omega^3} \tag{19}$$

The solution of (19) is $a(t)=a(0)e^{-ct/2}$ and $\Delta f = -5\alpha^2 a(0)^2 / 24\pi\Omega^3 e^{-ct}$. The amplitude of the impulse response is exponential decay, which is the same as the linear analysis result. The frequency of impulse response deviates from the natural frequency, which is different from the linear analysis result. Moreover, the frequency deviation is related to the FO amplitude. The greater the FO amplitude is, the larger the frequency deviation will be.

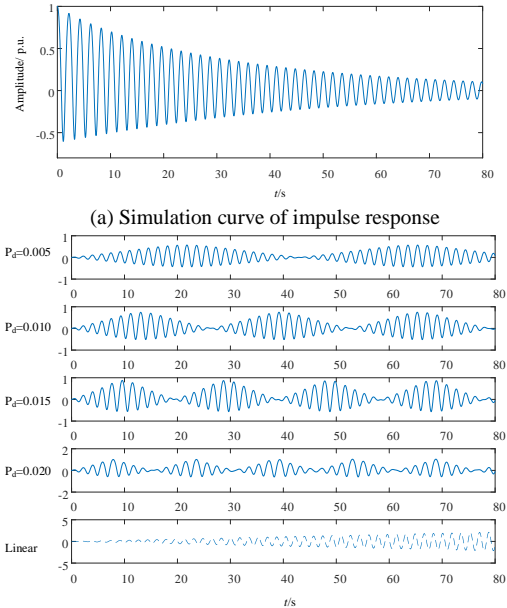

(a) Simulation curve of impulse response

(b) Simulation curve of resonance response at nature frequency

Fig 10. Simulation curve under frequency deviation.

Fig. 10 illustrates the simulation results under frequency deviation. It can be seen from Fig. 10(a) that the oscillation frequency of impulse response is 0.48Hz at initial and 0.57Hz at the end, which means that large amplitudes lead to more negative frequency deviations. Also the frequency deviations can be observed in Fig. 10(b), where $D=0$. The beat frequency rises as $P_d$ increases, which means the frequency deviations is enlarged when increasing the amplitude of FOs.

## V. CONCLUSION

This paper theoretically analyzed the explicit expression of the forced oscillation in the power system considering the nonlinear part and obtained the resonance curve. And theoretically, frequency deviation caused by nonlinearity is explained. Significant conclusions drawn from our analysis can be summarized as the following three points:

1) The FOs under linear and nonlinear are very different, especially when the FO amplitude is high.

2) The frequency deviations and jump phenomenon are observed in nonlinear FOs of power systems, which are their distinct characteristics compared with linear FOs.

3) The nonlinearity of power systems and large amplitude of disturbance cause stronger frequency deviations. And damping and inertia will suppress the frequency deviation .

REFERENCES

[1] Magdy, M.A. and F. Coowar, Frequency domain analysis of power system forced oscillations. IEE proceedings. Part C, Generation, transmission and distribution, 1990. 137(4): p. 261.
[2] Myers, R.B. and D.J. Trudnowski, Effects of forced oscillations on spectral-based mode-shape estimation, in 2013 IEEE Power & Energy Society General Meeting. 2013. p. 1-6.
[3] Y. Xu, Z. Gu, K. Sun and X. Xu, "Understanding a Type of Forced Oscillation Caused by Steam-Turbine Governors," in IEEE Transactions on Energy Conversion, vol. 35, no. 3, pp. 1719-1722, Sept. 2020, doi: 10.1109/TEC.2020.2995073.
[4] D. Wu, P. Vorobev, S. C. Chevalier and K. Turitsyn, "Modulated Oscillations of Synchronous Machine Nonlinear Dynamics With Saturation," in IEEE Transactions on Power Systems, vol. 35, no. 4, pp. 2915-2925, July 2020, doi: 10.1109/TPWRS.2019.2958707.
[5] Ye, H., et al., Analysis and Detection of Forced Oscillation in Power System. IEEE Transactions on Power Systems,2017. 32; 32(2; 2): p. 1149-1160.
[6] S. A. N. Sarmadi and V. Venkatasubramanian, "Inter-Area Resonance in Power Systems From Forced Oscillations," in IEEE Transactions on Power Systems, vol. 31, no. 1, pp. 378-386, Jan. 2016, doi: 10.1109/TPWRS.2015.2400133.
[7] L. Chen, Y. Min and W. Hu, "An energy-based method for location of power system oscillation source," in IEEE Transactions on Power Systems, vol. 28, no. 2, pp. 828-836, May 2013, doi: 10.1109/TPWRS.2012.2211627.
[8] Z. Wang and Q. Huang, "A Closed Normal Form Solution Under Near-Resonant Modal Interaction in Power Systems," in IEEE Transactions on Power Systems, vol. 32, no. 6, pp. 4570-4578, Nov. 2017, doi: 10.1109/TPWRS.2017.2679121.
[9] N. S. Ugwuanyi, X. Kestelyn, O. Thomas, B. Marinescu and A. R. Messina, "A New Fast Track to Nonlinear Modal Analysis of Power System Using Normal Form," in IEEE Transactions on Power Systems, vol. 35, no. 4, pp. 3247-3257, July 2020, doi: 10.1109/TPWRS.2020.2967452.
[10] Q Liu, Y Xue, G Chen ,"Oscillation Analysis Based on Trajectory Modes Decoupled in Space and Mode-energy-sequence Part Three Time-varying Resolution Thinning from Swing to Time Section", in Automation of Electric Power Systems, Vol.43 No.14, pp. 1-8, July 2019, DOI: 10.7500/AEPS20190430036.